\documentclass[twocolumn,prb,showpacs,amsmath,amssymb,floatfix]{revtex4}
\usepackage{graphicx}
\usepackage{ulem}
\newcommand{\br}{\mathbf{r}}

\newcommand{\bk}{\mathbf{k}}

\newcommand{\bq}{\mathbf{q}}

\newcommand{\bn}{\begin{equation}}
\newcommand{\ee}{\end{equation}}
\newcommand{\bga}{\begin{eqnarray}}
\newcommand{\eda}{\end{eqnarray}}
\newcommand{\half}{\frac{1}{2}}
\newcommand{\diff}{\text{d}}
\newcommand{\eps}{\epsilon}
\newcommand{\A}{\text{\AA}}

\newcommand{\chalmersMC}{$^1$Department of Microtechnology and Nanoscience, MC2,
Chalmers University of Technology,
SE-41296 G\"{o}teborg, Sweden}

\begin{document}
\title{An analysis of van der Waals density functional components: Binding and corrugation of benzene and C60 on  boron nitride and graphene}
\author{Kristian Berland}
\email{berland@chalmers.se}
\author{Per  Hyldgaard}
\email{hyldgaar@chalmers.se}
\affiliation{\chalmersMC}
\date{\today}
\begin{abstract}
The adsorption of benzene and C60 on graphene and boron nitride (BN) is studied using density functional theory with the non-local correlation functional vdW-DF.
By comparing these systems we can systematically 
investigate their adsorption nature and 
differences between the two functional versions vdW-DF1 and vdW-DF2.
The bigger size of the C60 molecule 
makes it bind stronger to the surface than benzene, yet the interface between the molecules and the sheets are similar in nature.  
The binding separation is more sensitive to the exchange variant used in vdW-DF than to the correlation version. This result is related to the exchange and correlation components of the potential energy curve (PEC). 
We show that a moderate dipole forms for C60 on graphene, unlike for the other adsorption systems. 
We find that the corrugation is very sensitive to the variant or version of vdW-DF used, in particular the exchange. Further, we show 
that this sensitivity arise indirectly through the shift in binding separation caused by changing vdW-DF variant. 
Based on our results, we suggest
a concerted theory-experiment approach to assess the 
 exchange and correlation contributions to physisorption.
Using DFT calculations, the corrugation can be linked to the optimal separation, allowing us to extract 
the exchange-correlation part of the adsorption energy. Molecules with same interfaces to the surface, but different geometries, can in turn cast light on the role of van der Waals forces. 
\end{abstract}
\maketitle

\section{Introduction}

The development of methods that include van der Waals forces in density functional theory (DFT) has 
made it possible to describe an abundance of important material systems ---
like  water,\cite{Water:Kelkkannen,water:dynamic} DNA,\cite{cooper:DNA} molecular crystals,\cite{nabok:245316,molcrys1,molcrys:piconj,molcrys2} and metal-organic frameworks\cite{MOFs,Storage:MOF} ---
at the electronic level even without use of empirical parameters. 
Such materials can be categorized as sparse matter,  
because of the essential role of the low (electron) density regions or ``voids'' in these materials. 
These regions make it paramount to go beyond semi-local approximations like the generalized-gradient approximation~(GGA)\cite{Langreth:WaveVector,rev:SpinDensity,PW86,B88,GGA:PW91,GGA:PBE,GGA:AM} and properly account for van der Waals interactions. 
A good description of sparse matter is important for developing new nanotechnology. The softer van der Waals bonds makes it easy for fragments to move,
like in the self assembly of molecules on surfaces~\cite{nanoLetter,benzCu} or for moving pieces of nano-mechanical devices.\cite{Capasso:nanomechanical} 

Sparse matter constitutes a research-frontier for DFT and DFT-based modelling.\cite{Klimes:Perspectives,Review:vdW}
Practical and robust schemes to describe sparse matter are all fairly recently developed and several competing and complimentary methods exist.
Two prominent approaches are vdW-corrected DFT
and the use of non-local correlation functionals. 
The first approach adds a semi-empirical pair-potential correction accounting for van der Waals forces on top of semi-local DFT.\cite{Wu:Yang,DFT-D,DFT-D2,DFT-D3} 
There exist many variation thereof and refinements, like adjusting the pair-potentials by using charge-density input.\cite{TS}
Non-local correlation functionals aim to stay fully within DFT keeping the density as the key variable. 
The van der Waals density functional~(vdW-DF) method\cite{Rydberg:Tractable,Rydberg:Layered,Dion:vdW,Lan:vdWApp,vdW:selfCons,Review:vdW} for approximating the exchange-correlation energy has led to a series of such functionals that lack empirical input. 
The vdW-DF1,\cite{Dion:vdW} introduced in 2004, and vdW-DF2,\cite{vdWDF2} introduced in 2012, emphasise a constraint-based design of the density fluctuation propagator,\cite{Lundqvist:I,Lundqvist:II,Lundqvist:III}  while the precursor for layered systems\cite{Rydberg:Tractable,Rydberg:Layered} emphasise an anisotropic screening account. 
The vdW-DFs vanish seamlessly in the uniform limit\cite{Dion:vdW}  and exhibits the correct $1/r^6$ asymptotic limit 
between molecular dimers.\cite{Dion:vdW} 
The prediction of various sparse matter methods can differ quite much. 

Hand in hand with the development and refinements of sparse matter methods, it is essential to identify experiments and quantum-chemistry calculations that serves to test and elucidate these methods for various physical effects and across length scales.\cite{Bjork:QW,Perdew:C60,Becke:HigherOrder,molcrys1,Barask1,Perdew:Asymptotic}
One approach is to construct a representative data set that account for many physical effects,\cite{vdW:ready,S22,S22:5,S66} like the S22 data set\cite{S22} 
or the more recent and larger S66.\cite{S66} 
The average performance of methods can then be compared to each other. 
Another approach is to identify particular systems and associated experiments that reveal several properties of sparse matter binding.\cite{Dobson:DispersionEnergies}
An example is to calculate potential energy curves~(PEC) of H$_2$ on Cu(111) and compare with the experimental ones constructed from backscattering experiments.\cite{H2onCu111,H2onCu111_B}
Since the balance between the different contributions to PEC change with separation, much information can be extracted. 

\begin{figure}[t]
  \centering
  \includegraphics[width=0.5\textwidth]{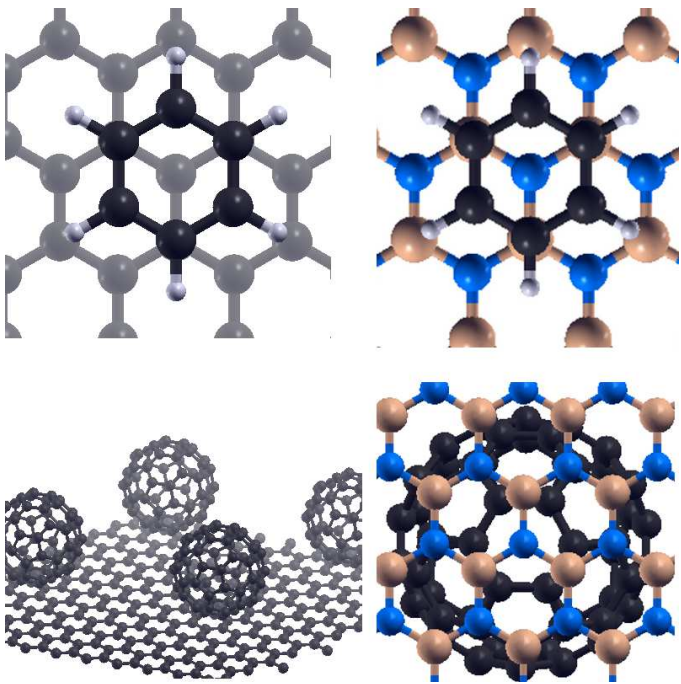}  
  \caption{Adsorption systems studied. Upper left: benzene on graphene on {\it top} site. Upper right: benzene on BN on {\it top-N} site. Lower left: Repeated supercells of C60 on graphene. Lower right: C60 on BN on {\it top-N-rot} site (viewed from below BN sheet).}
  \label{fig:Ads_system}
\end{figure}

In this paper, we show that the details of adsorption, especially the predicted corrugation for adsorbate dynamics is a discriminator for assessing the balance between exchange and correlation components in a non-local functional design. 
We substantiate our conclusion by comparing the results of calculations using vdW-DF1 and vdW-DF2 and associated exchange choices for a group of adsorbate systems that are closely related, yet the nature of the binding differs in a well-defined manner. 

We study the adsorption of benzene and C60 on graphene and BN. We show how these systems allow us to span, in a systematic manner, a wide range of different contributions to sparse matter binding. 
First, the interface between a benzene and
graphene (or BN) is very similar to that of C60 on graphene (or BN), both in terms of contact area and character of the local bonds. 
Yet C60 is bigger and the difference in adsorption is mostly induced by the increased van der Waals attraction of the C60 molecule. 
Second, boron nitride and graphene are both flat one-atom layer thick sheets with eight valence electrons per unit cell, but whereas graphene is a semi-metal, BN is an insulator. 
The BN sheet also has a somewhat different charge-density landscape than the graphene sheet.

We show that the corrugation is very sensitive to the binding separation and therefore very sensitive to the version of 
correlation and exchange in particular used in vdW-DF. 
Yet at a given separation to the surface, the corrugation is largely independent of the functional version used. 
The shorter binding separation for C60 adsorbants compared to benzene increase corrugation in the same indirect manner. 

Based on our results and analysis, we suggest a combined theory-experiment approach that can lead to insight into the nature of sparse matter binding and the magnitude of the different terms in the exchange-correlation functional.

The plan of the paper is as follows: After the introduction we summarize vdW-DF method and functional versions and variants and outline computational details specific to this study. 
Section~\ref{sec:Adsorption} presents our results for adsorption of benzene and C60 on graphene and boron nitride. 
In section~\ref{sec:Analysis} we analyze the various components to this binding, both the importance of GGA exchange choice and vdW-DF correlation version. We also analyze the role of different density separations and density regimes in the non-local correlation of vdW-DF.
Section~\ref{sec:corrugation} compares the corrugation of the different systems and for the different functionals. 
Section~\ref{sec:discussion} discuss the possibility of using the selected systems to assess the 
quality of non-local correlation functionals and sparse-matter methods in general. 
Finally, we summarize our conclusions. 

\section{vdW-DF calculations}

\begin{figure}
  \includegraphics[width=0.5\textwidth]{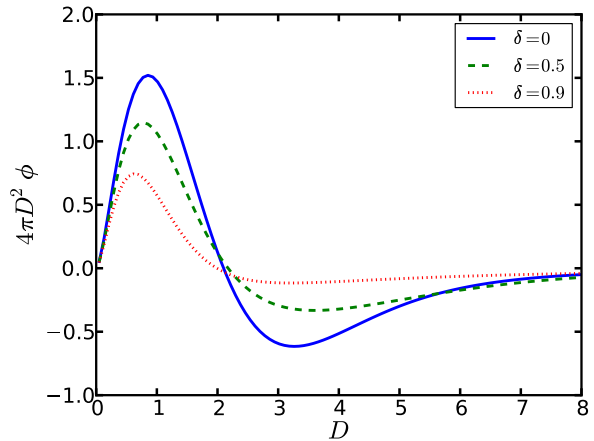}
  \caption{The vdW-DF kernel. Figure similar to that of Dion~et.al.~\cite{Dion:vdW} 
}
  \label{fig:kernel}
\end{figure}

The vdW-DF method is designed to handle both short-range covalent bonds and long-range van der Waals interactions without introducing damping functions or empiricism. 
Its exchange-correlation functional combines exchange  at the GGA level, $E_{\rm x}^{\rm vdWDF} = E_x^{\rm GGA}$
with the combination of correlation in the local density approximation (LDA) and non-local correlation: $E^{\rm vdWDF}_{\rm c} = E^{\rm LDA}_{\rm c} + E^{\rm nl}_{\rm c}$.
The non-local correlation is 
responsible for the van der Waals interaction.

The non-local correlation term vanishes in the uniform limit and takes the form of a 
six dimensional integral 
\begin{equation} 
  E^{\rm nl}_c[n]= \half \int\diff^3 \br \int\diff^3 \br'\, n(\br) \phi[n](\br,\br') n(\br')\,.\label{eq:Enl}
\end{equation}
The kernel can be expressed in terms of a universal kernel $\phi(\br,\br') = \phi_1(d,d')$ that depends on  
two dimensionless length scales $d= q_0(\br_1)r_{12} $ and $d'= q_0(\br_2) r_{12}$
where
$r_{12} =  |\br_1 - \br_2|$ and $q_0$ is a modulation of the local Fermi vector $k_{\rm F}(n)$ that account for the local response. 
By depending on gradients as swell as local density, this local response 
reflects a broader dependence on the  density variation through a plasmon-pole approximation.\cite{Dion:vdW}

The universal kernel $\phi_1$ kernel can also be tabulated in terms of $D=(d+d')/2$ and $\delta = (d-d')/(d+d')$. 
Fig.~\ref{fig:kernel}  shows the kernel as a function of $D$ for three selected $\delta$. 
The $\delta$ parameter is a measure of how different the local response of the two density regions are.

The GGA exchange energy can be expressed as a modulation $F_x$ of the LDA exchange energy $\eps_{\rm x}^{\rm LDA}(n(\br))$, as follows
\begin{equation}
  E_{\rm x}^{\rm GGA} = \int \diff^3 \br \, n(\br) \eps_{\rm x}^{\rm LDA}(n(\br)) F_{\rm x}(s) \,.
  \label{eq:Ex}
\end{equation}
Here $s = (|\nabla n|/2n) / k_{\rm F}$ is a measure of how rapidly the density is changing compared to length scale set by the local Fermi wave vector. In the small $s$ limit, the modulations of the GGAs go like  $F_x(s) -1 \sim s^2$, while in the large $s$ limit they differ significantly both quantitatively and qualitatively.

\subsection{Functional versions}
\label{sec:func_version}

Because the nature of the adsorption systems considered here differ in a systematic manner, 
it becomes interesting to study how the predictions of vdW-DF changes with correlation version and choice of exchange partner. 
For vdW-DF1 correlation\cite{Dion:vdW}~(the version of 2004), we will in, in addition to the canonical choice of revPBE, use C09,\cite{Cooper2009} PW86r,\cite{ExcEnergy,PW86} and optPBE\cite{Klimes} as exchange partner. 
For vdW-DF2 correlation,\cite{vdWDF2} we use C09, in addition to the canonical choice of PW86r. 

Standard vdW-DF1 with the revPBE~\cite{GGA:revPBE} exchange functional as companion to its non-local correlation
 have shown 
an impressive robustness, giving good results for binding between molecules of various sizes and properties,\cite{Dion:vdW,vdW:selfCons,Review:vdW,puzder:benzene,cooper:DNA,DimersNew,Water:Kelkkannen,water:dynamic,Ice:vdWDF,molcrys1,molcrys2} adsorption onto insulating surfaces,\cite{bz:SnO2,Moses:BenzeneMos2,butane:Cu(111),Karen:BenzeneSilicon} graphene,\cite{Svetla:BenzeneGraphite,adenineGraphene,Londero:alkanes} and metals,\cite{Sony:Cu110,benzCu,H2onCu111,H2onCu111_B,RPBE:Au(111),effectiveElastic,Neaton:Ads_metalSurface} and between various layered compounds.\cite{Layered:Risto,Londero1} 
Yet, its accuracy for sparse matter
does not provide the full accuracy of traditional DFT calculations for typical covalently bonded syste, where the generalized-gradient approximation excels.
Some of this inaccuracy can be attributed to the overly repulsive
revPBE\cite{GGA:revPBE} exchange functional causing overestimated separations. This exchange functional was chosen because many other standard choices like PBE~\cite{GGA:PBE} tend to produce spurious 
binding arising from exchange effects.\cite{Wu_Scoles:Towards_Extending,Kannemann_Becke,ExcEnergy}

Some sparse matter systems are also particularly challenging, like those with weak chemisorption. These systems are highly sensitive to the balance between the different terms in the exchange-correlation functional.\cite{weak:Andre,C60onAu(111),graphene:surfaces,Stepped:Donatio,Blugel:Cu110,graphene:Ir,graphene:Ir2}
The overestimation of binding separations in vdW-DF can make it miss a van der Waals-induced charge transfer, a signature of weak chemisorption. 

Refinements of the vdW-DF1 functional have been suggested, in particular it has been suggested to replace its exchange functional with a less repulsive one. 
Cooper~\cite{Cooper2009} designed an exchange functional C09 based on different fundamental constraints than revPBE, 
following the principles that underpins the introduction of the PBEsol functional.\cite{PBEsol} C09 is less repulsive than revPBE, but becomes similar in the large $s$-limit, thereby minimizing contributions to unphyscial exchange binding. 

Klime\v{s} and coworkers\cite{Klimes} reparameterized a set of exchange functionals by optimizing their performance 
to the S22 data set.\cite{S22}
One of these, the optPBE functional, corresponds to a reparameterization of PBE and is therefore somewhat similar to revPBE. This functional will be included in our study. 

Recently, a new version of vdW-DF called vdW-DF2 has been developed.\cite{vdWDF2} 
This functional sets the $q_0(\br)$ parameter by using a different description of the plasmon
that is suggested by an analysis of the exchange-variation in the high-density limit.\cite{Schwinger1,Schwinger2,Elliott_Burke:B88}
The updated plasmon description makes vdW-DF2 less sensitive to low-density regions and generally reduces the non-local correlation energy. 
In addition to adjusting the version of correlation, this functional updates the account of exchange by relying on a refitted version of PW86, here labeled PW86r.\cite{PW86,ExcEnergy}
This exchange functional agree well with the exchange of Hartre-Fock calculations and avoids spurious exchange binding.\cite{Harris,ExcEnergy,Wu_Scoles:Towards_Extending}
This functional is also less repulsive than revPBE at binding separations. 

In almost every case, the modifications of vdW-DF description reduces binding separations compared to the results of vdW-DF1. 
For the S22 data set,\cite{S22} vdW-DF1 with C09 exchange and vdW-DF2 produces good results; 
 unsurprisingly so does using vdW-DF1 with optPBE, since it is was fitted to this data set. 
When comparing the versions for a broader class of systems, we find that no version clearly outperforms the others. 

To exemplify this observation, 
vdW-DF2 stands out by predicting a potential energy curve for H$_2$ on Cu(111)  that compares well with that generated from experimental data,\cite{H2onCu111,H2onCu111_B} 
both in terms of optimal separation and adsorption energy. However it underestimates the asymptotic form. 
For C60 crystals,\cite{molcrys2} vdW-DF1 predicts a cohesive energy 1.6~eV in good agreement with experimental measured interval 1.6-1.9. 
vdW-DF1 using C09 exchange functional predicts an excellent lattice constant and a cohesive energy of about 2~eV. In contrast vdW-DF2 predicts a cohesive energy of merely 1.3~eV.
For adsorption of C60 on Au(111), the combination of vdW-DF2 and C09 has been found to yield the best results.\cite{C60onAu(111)}
For this system, neither vdW-DF1 nor vdW-DF2 are able to predict the right balance between repulsive and attractive terms that results in the experimental finding of a weak chemisorption.\cite{weak:Andre}
We find that vdW-DF1 predicts the best binding energies for both benzene and C60 on graphene. 

Bj\"orkman and coworkers recently asked the question ``are we van der Waals ready?'' in a large survey of the ability 
of electronic structure methods, including vdW-DF1 and vdW-DF2, to accurately describe binding of layered systems.\cite{vdW:ready} None of the DFT functional methods they tested are able to consistently predict accurate binding energies, separations, and elastic constants. 
For instance both vdW-DF1 and vdW-DF2 in general overestimates interlayer separations. 

To get a better understanding of why different functionals prevail for different systems,  a more systematical analysis of their binding mechanisms are needed. 
A suggestion to test performance by comparing with experiments that reveal many interaction properties for the same one system was launched in Refs.~\onlinecite{H2onCu111,H2onCu111_B}
Here, we seek to map out how both the exchange and correlation functionals contribute to PECs for related molecules of different sizes and based on this analysis propose an additional adsorption-based strategy to refine functional testing.
Some of the analysis approaches taken here can be extended to other van der Waals bonded systems, or directly applied to other non-local correlation functionals like those of Vydrov and Voorhis.\cite{MIT1,MIT2,MIT3}

\subsection{Computational method}

The vdW-DF adsorption study presented here relies on a post-GGA procedure similar to many previous vdW-DF studies: the charge density is first generated in a self-consistent PBE calculation with ultrasoft pseudopotentials using the \textsc{Dacapo} software package;\cite{DACAPO} next, in a post-processing procedure, we calculate the semi-local exchange at the GGA level and non-local vdW-DF correlation. These terms replace those of the self-consistent calculation. 
The non-local correlation part is calculated with an in-house real-space code presented and discussed in Ref.~\onlinecite{molcrys2}. 

This study is similar to previous vdW-DF adsorption studies detailed in Refs.~\onlinecite{benzCu,adenineGraphene,graphane} and resolves computational issues like grid sensitivity and handling of supercells in the same manner. 
Here, the plane-wave cutoff is set to 500~eV and a $\bk$-sampling of $2\times2\times1$ is used. 
The supercell consists of $6\times6$ graphene (boron nitride) primitive unit cells. This size makes inter-adsorbate interaction negligible even for the case of C60 adsorbates (see lower left of Fig.~\ref{fig:Ads_system}), since the non-local interaction between adsorbates is explicitly corrected for. 

It is useful to consider the part of the total energy in vdW-DF that contains one-particle kinetic, electrostatic, local correlation, and exchange $E_0$ separately from the remaining non-local correlation $E^{\rm nl}_{\rm c}$. 
In a purely van der Waals bonded system, the former part typically gives rise to a repulsive contribution to the potential energy curve~(PEC), while the non-local correlation part, an attractive. 
In calculating the contributions to the PEC between a molecule and surface, we here calculate the former by subtracting off the energy off a reference system where the molecule is far from the surface 
\begin{equation}
  \Delta E_0  = E_0(h) - E_0(h_\infty)\,;
  \label{eq:E0}
\end{equation}
in practice $h_\infty =9$~\AA\, is sufficient.
While for the contribution arising from non-local correlation, which have longer range, we subtract off the energy of isolated fragments\cite{C6:Kleis,molcrys2}
\begin{equation}
  \Delta E^{\rm nl}_c  = E^{\rm nl}_{\rm c,full}(h) - E^{\rm nl}_{\rm c,sheet} - E^{\rm nl}_{\rm c,molecule}\label{eq:Delta_Enl}\,.
\end{equation}

\section{Adsorption results}
\label{sec:Adsorption}

We describe the adsorption results of our vdW-DF study of adsorption of benzene and C60 on graphene and boron nitride.
We present the binding energies and sheet-to-molecule separation for the different systems and for the different functionals versions.
We also calculate the dipole induced by adsorption and show that only C60-on-graphene induces a significant dipole at binding separations. 


\subsection{Binding energy and separation}
\begin{figure}
  \centering
  \includegraphics[width=0.5\textwidth]{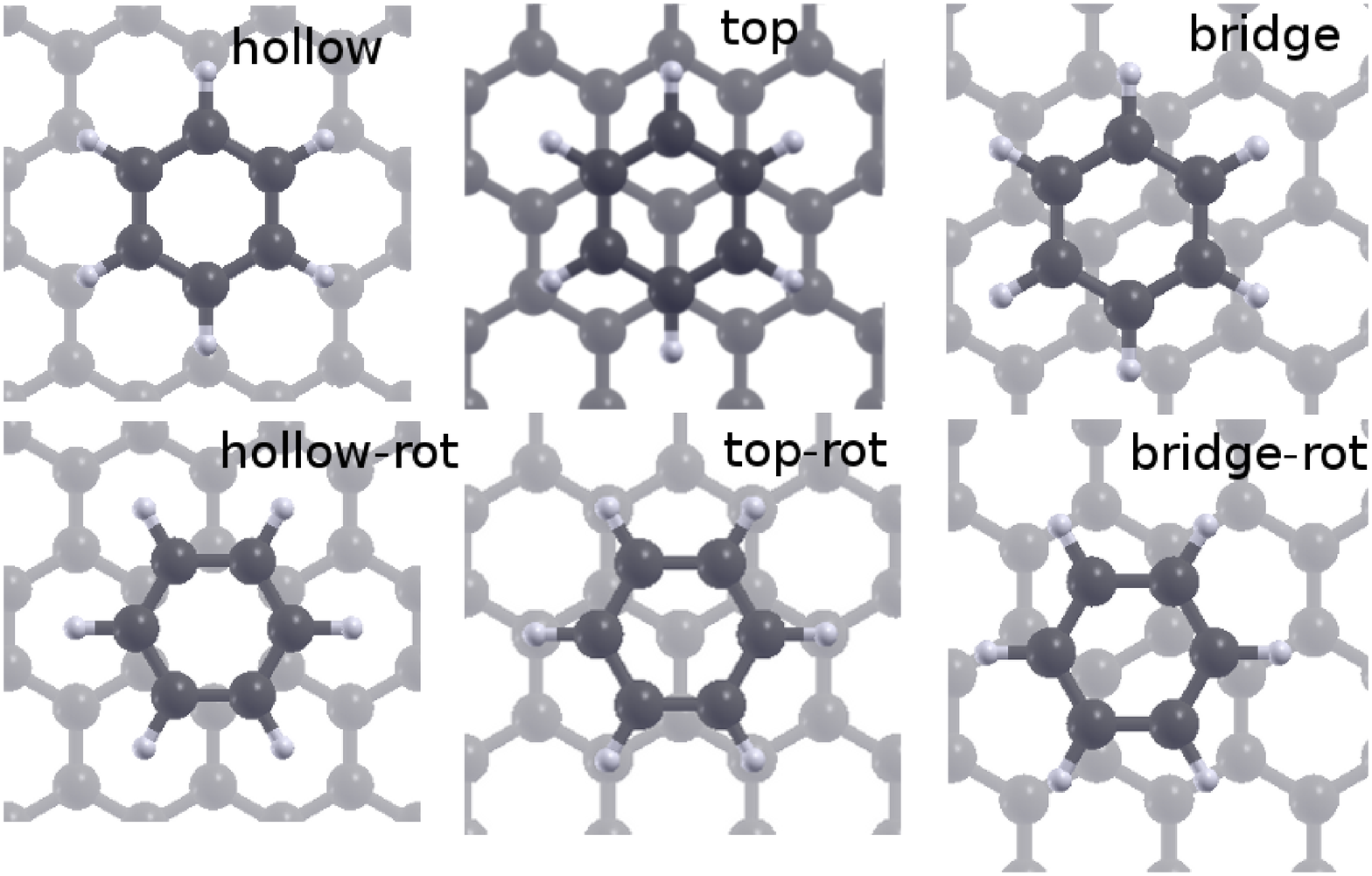}
  \caption{Adsorption sites considered for benzene on graphene. The -rot sites are rotated $30^{\circ}$. The top site is the optimal one.}
  \label{fig:site_bz_g}
\end{figure}

To determine the optimal site and corresponding binding energy and separation, we 
optimize the separation to the sheet for each high-symmetry site. Fig.~\ref{fig:site_bz_g} shows the six high symmetry sites considered for benzene on graphene. The molecule is assumed to lay flat on the sheet. 
On boron nitride, we consider eight high symmetry sites, since the center ring can be on-top of a nitrogen ({\it top-N/top-N-rot}) and on-top of a boron ({\it top-B/top-B-rot}). 
For C60 we consider both configurations where the carbon hexagon or pentagon face the sheet and find that the hexagon is the preferred one. 
For benzene, the optimal site is the {\it top} on graphene and  {\it top-N} on boron nitride.
The optimal hexagonal configuration are rotated $30^\circ$ degrees about the surface normal $\hat{z}$ for C60 compared to the optimal one for benzene ({\it top-rot/top-N-rot}). 
The same optimal sites were predicted by the considered versions and variants of vdW-DF.

\begin{figure}
\includegraphics[width=0.5\textwidth]{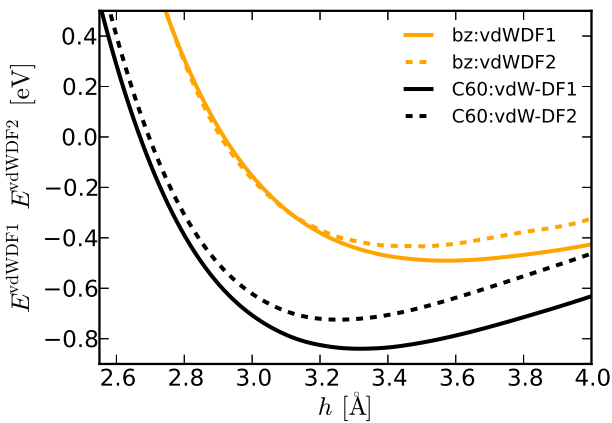}
\caption{The potential energy curves (PEC) for benzene~(yellow curves) and C60 (black) for adsorption on graphene as obtained with vdW-DF1~(full) and vdW-DF2~(dashed).}
\label{fig:binding}
\end{figure}

\begin{table}[h]
  \caption{Adsorption energy $E_{\rm ads}$ and separation $h$ between the molecule and graphene (closest atom) as obtained for different variants of vdW-DF.
  Energies are compared to the experimental for adsorption ongraphene.}
\begin{ruledtabular}
\begin{tabular}{lll}
  & \begin{tabular}{l}  benzene on graphene\\ \hline \end{tabular} &  \begin{tabular}{l}  benzene on BN \\ \hline \end{tabular} \vspace{0.1cm} \\
  \begin{tabular}{l}
Functional    
\vspace{0.1cm}\\
{\bf vdW-DF1}      \\
vdW-DF1(C09) \\
vdW-DF1(optPBE)\\
vdW-DF1(PW86r) \vspace{0.1cm}\\
{\bf vdW-DF2} \\
vdW-DF2(C09) \vspace{0.1cm} \\
Exp.    
\end{tabular}
& 
\begin{tabular}{ll}
 $h\,[\text{\AA}]$ & $E_{\text{ads}}\,[\text{eV}]$\vspace{0.1cm}
 \\
  3.6    & 0.49\\
 3.25   &  0.59 \\
 3.4  &0 59  \\
 3.35   & 0.66 \vspace{0.1cm}\\
   3.5     &  0.43  \\
 3.4 & 0.34  \vspace{0.1cm} \\
  - & 0.50$\pm$0.08\footnotemark[1] 
 \end{tabular}
 &
\begin{tabular}{ll}
 $h\,[\text{\AA}]$ & $E_{\text{ads}}\,[\text{eV}]$ \vspace{0.1cm}\\
 3.5  & 0.48   \\
  3.2 & 0.59    \\
 3.35 & 0.58    \\ 
 3.35   & 0.66  \vspace{0.1cm} \\
  3.4 & 0.43  \\
3.3  &  0.34 \vspace{0.1cm}\\
- & - \\
\end{tabular}\vspace{0.2cm}
\\
  & \begin{tabular}{l}  C60 on graphene\\ \hline \end{tabular} &  \begin{tabular}{l}  C60 on BN \\ \hline \end{tabular} \vspace{0.1cm} \\
\begin{tabular}{l}
Functional    
\vspace{0.1cm}\\
{\bf vdW-DF1}  \\ 
vdW-DF1(C09) \\
vdW-DF1(optPBE)\\ 
vdW-DF1(PW86r)  \vspace{0.1cm}\\
{\bf vdW-DF2}    \\
vdW-DF2(C09)  \\
Exp.   \\
\end{tabular}
&
\begin{tabular}{ll}
 $h\,[\text{\AA}]$ & $E_{\text{ads}}\,[\text{eV}]$\vspace{0.1cm}  \\
 3.3   &  0.85   \\
3.0    &  1.06  \\
 3.15 &  1.01  \\
 3.15   & 1.10   \vspace{0.1cm}\\
 3.25    & 0.72    \\
 3.1  & 0.65  \vspace{0.1cm} \\
  - &  0.85\footnotemark[2] 
 \end{tabular}
 &
\begin{tabular}{ll}
 $h\,[\text{\AA}]$ & $E_{\text{ads}}\,[\text{eV}]$ \vspace{0.1cm}\\
  3.3    &  0.83     \\
  3.0   & 1.04  \\
3.15 &   0.99  \\
  3.15   &   1.07   \vspace{0.1cm}\\
 3.25 &    0.69  \\
  3.1  & 0.63  \vspace{0.1cm} \\
   - & - \\
\end{tabular}
\end{tabular}
\footnotemark[1]{Ref.~\onlinecite{PolyAromitc:HC}}
\footnotemark[2]{Ref.~\onlinecite{C60graphene}}
\end{ruledtabular}
\label{tab:energies}
\end{table}

Figure~\ref{fig:binding} shows PECs for benzene and C60 on graphene as calculated in vdW-DF1 and vdW-DF2. 
Table~\ref{tab:energies} displays the adsorption energies and the optimal molecule-to-sheet separations for the optimal sites
including adsorption on BN and use of non-canonical exchange partners. 

The benzene adsorption energies on graphene almost match those on BN, while C60 binds slightly stronger to graphene.
The optimal benzene-on-graphene separation is slightly larger than the benzene-on-BN, while the C60-on-graphene and C60-on-BN separations are the same. 
That benzene and C60 does not follow identical trends could stem from a moderate charge transfer between graphene and C60, compared to a much smaller one for the C60 on BN. 
The difference in charge transfer is discussed in the next subsection. 


vdW-DF1 predicts adsorption energies very close to the experimental for both benzene and C60 on graphene. 
vdW-DF2 somewhat underestimates the experimental value. 
vdW-DF1 correlation combined with the non-canonical exchange variants considered here all result in overestimated adsorption energies, while vdW-DF2 with C09 exchange underestimates the adsorption energies. 

Both vdW-DF1 and vdW-DF2 describe well the expected increased van der Waals attraction for the larger system. vdW-DF1 predicts that C60 binds 73\% stronger than benzene on graphene,
vdW-DF2 predicts 67\%, and
the experimental number is 70\%.
Note though, the specific factor change if a different exchange functional is used. 

We are unaware of explicit experimental adsorption energies for benzene and C60 on BN. 
Reinke and coworkers did find that those C60 molecules 
not associated with defects desorbed at slightly lower temperature on boron nitride than on graphene.~\cite{reinke:B60_bn_graphene} This observation is consistent with our results. 
Binding separations are unavailable for any of the considered systems, complicating a clear assessment of the functionals.
vdW-DF1 typically overestimates binding separations by 
about 0.2-0.3~\AA\, for purely van der Waals bonded systems.\cite{molcrys1,molcrys2,Review:vdW} 
We find that the other variants reduce the separations, in particular those using C09 exchange. 

Caciuc~and coworkers~\cite{graphene-boron} 
also calculated benzene adsorption on graphene and BN both 
for vdW-DF1 and vdW-DF2.
Our results agree well with theirs, with the exception their separations are about 0.1~\AA~shorter for benzene on BN than ours. 
Using with semi-empirical DFT-D2,\cite{DFT-D2} they found the same optimal sites as us.
The fact that their study is based on a projector-augemented waves implementation in \textsc{VASP},\cite{VASP-PAW} while ours rely on ultasoft pseudopotential and \textsc{DACAPO},\cite{DACAPO}
strengthens our confidence in the overall precision of 
both our and their study. 
The adsorption results for benzene on graphene 
also closely match the results in the seminal study by Chakarova-K\"{a}ck and coworkers.\cite{Svetla:BenzeneGraphite}

\subsection{Induced dipole}

\begin{figure}
  \centering
  \includegraphics[width=0.3\textwidth]{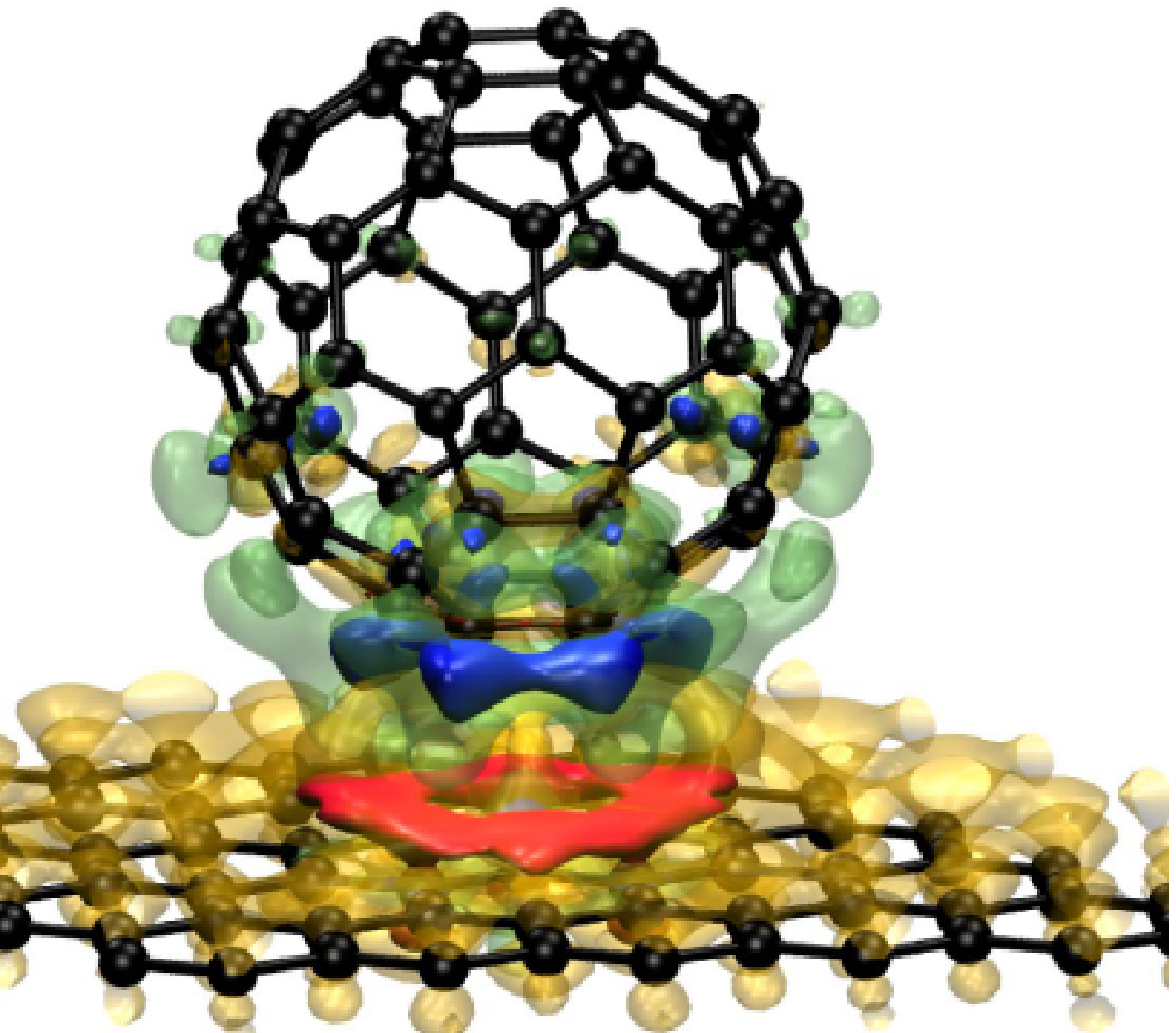}  
  \includegraphics[width=0.5\textwidth]{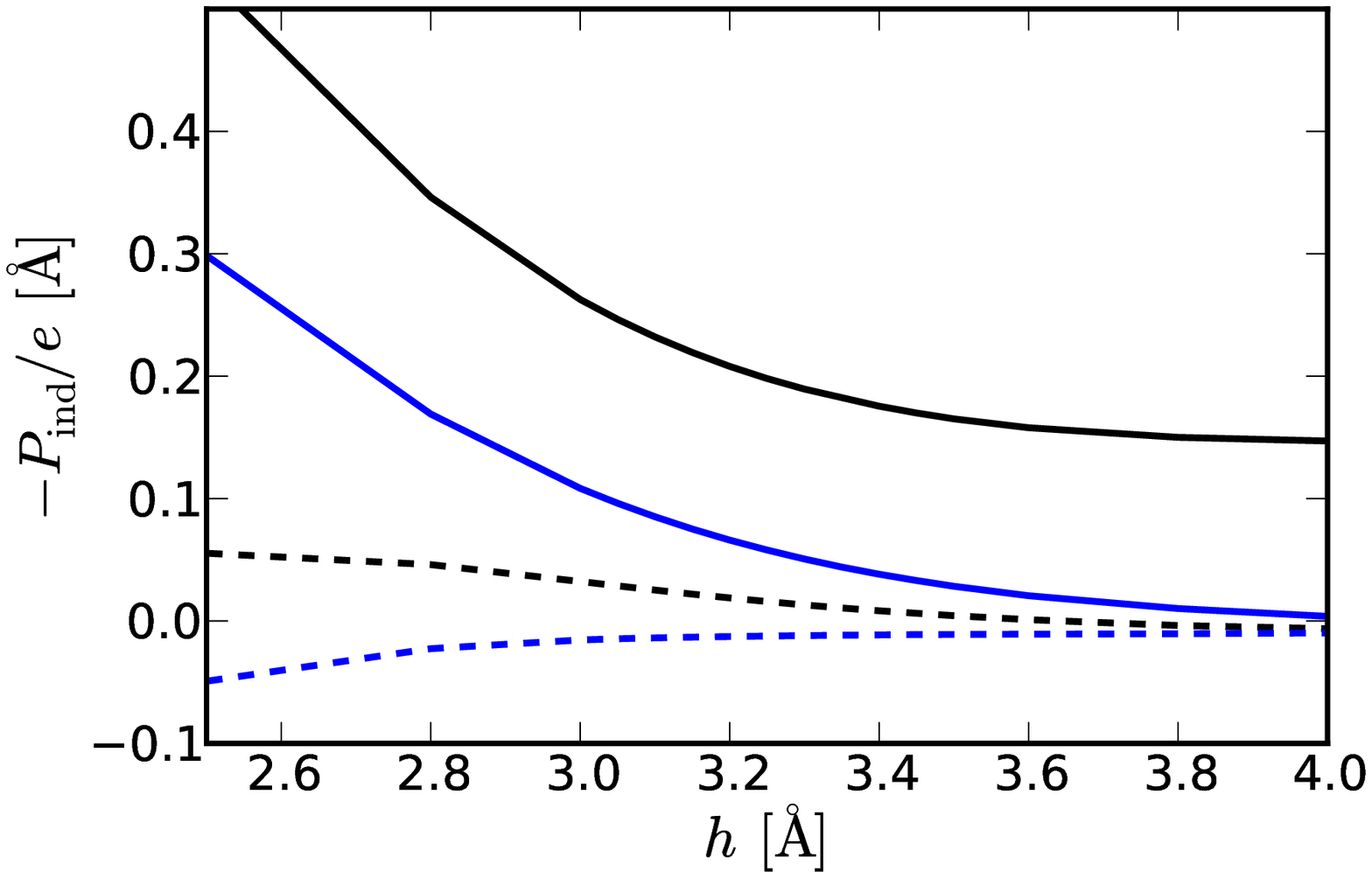}  
  \caption{
  Upper panel: Charge transfer isosurfaces for C60 on graphene $\Delta \rho(r)$. Blue is negative, red is positive. 
 Lower panel: Dipole induced as a function molecule-to-sheet separation. Upper full black curve is for C60 on graphene, lower full blue is for C60 on boron nitride. Upper dashed is for benzene on graphene and lower is for benzene on BN.}
  \label{fig:dipole}
\end{figure}

The upper panel of Fig.~\ref{fig:dipole} visualizes the charge transfer density induced by the adsorption of C60 onto graphene, as given by $\Delta \rho(\br) = -e \Delta n(\br) =  \rho_{\rm C60-graphene}(\br) - \rho_{\rm graphene}(\br) - \rho_{\rm C60}(\br)$. 
In the DFT calculations, grid sensitivity was minimized by using the same the same atomic coordinates relative to the underlying grid for the full systems as for the separate fragments. 
The induced charge transfer density has a rich dipolar structure.
The magnitude of this dipole depends on the separation between the molecule and the sheet. 

The lower panel of Fig.~\ref{fig:dipole} shows the induced dipole as function of the benzene-to-sheet (dashed) and C60-to-sheet separation (full) for both graphene (black) and BN (blue) sheets. 
At typical binding separations and beyond, only C60-on-graphene produces a non-negligible dipole. 
The dipole that forms between C60 on BN is tiny, yet if the molecule is pushed towards the surface, the magnitude of the dipole increases; while for benzene adsorbants, the magnitude of the dipole barely changes with separation.

To get a crude estimate for the total charge transfer from C60 to graphene, we first project the charge transfer density $\Delta \rho(\br)$ onto the z-axis  $\Delta \rho(z)$. We next identify the position $z_0$ where $\Delta \rho(z_0)=0$ that lies between the adsorbant and the sheet $\Delta \rho(z_0)=0$. 
Finally, we integrate over the projected charge transfer density smaller than $z_0$: $\Delta Q = \int^{z_0}_{-\infty}\diff z\, \Delta \rho(z)$.
The estimated charge transfer, of about $\Delta Q \approx  0.04 e$, is quite insensitive to binding separation, just like the induced polarization $P_{\rm ind} = \int \diff z\, z \Delta \rho(z) $ (as shown in the lower panel).

The estimated charge transfer is moderate compared to 0.3~e found for tetrafluoro-tetracyanoquinodimethane $\rm C_{12}(NF)_4$ on graphene.\cite{ptype:graphene}
This molecule, basically a hexagonal carbon ring with two arms of (NC)$_2$ and four fluorine (F) in place of H, has been proposed as a p-dopant on graphene. 
That this systems results in a significant charge transfer is not surprising considering the electronegativity of fluorine.
Since we find some charge transfer for C60 and none for benzene, we speculate that functionalized C60 might be a better candidate than functionalized benzene for doping graphene.

\section{Analysis of exchange and correlation part of vdW-DF binding}

We analyze the role of exchange and correlation in the vdW-DF account of adsorption of benzene and C60 on graphene.
This analysis benefits from comparing these two molecules, since they have a similar interface to the sheets, but different sizes. 
This property is first used to illustrate the inherit failure of local approximations to the exchange-correlation functional, as they are unable to capture the increased attraction of C60. 
Next, we examine the role of different exchange and correlation components to the potential energy curves of benzene and analyze the adsorption energy trends in light of these. 
We also confirm that the bulk of the difference between the PECs curves of benzene and C60 adsorption arise from non-local correlation.  
A more detailed analysis of the non-local correlation of vdW-DF follows including the role of density separations and density regimes.

\label{sec:Analysis}

\subsection{Failure of local approximations}

\begin{figure}
  \includegraphics[width=8cm]{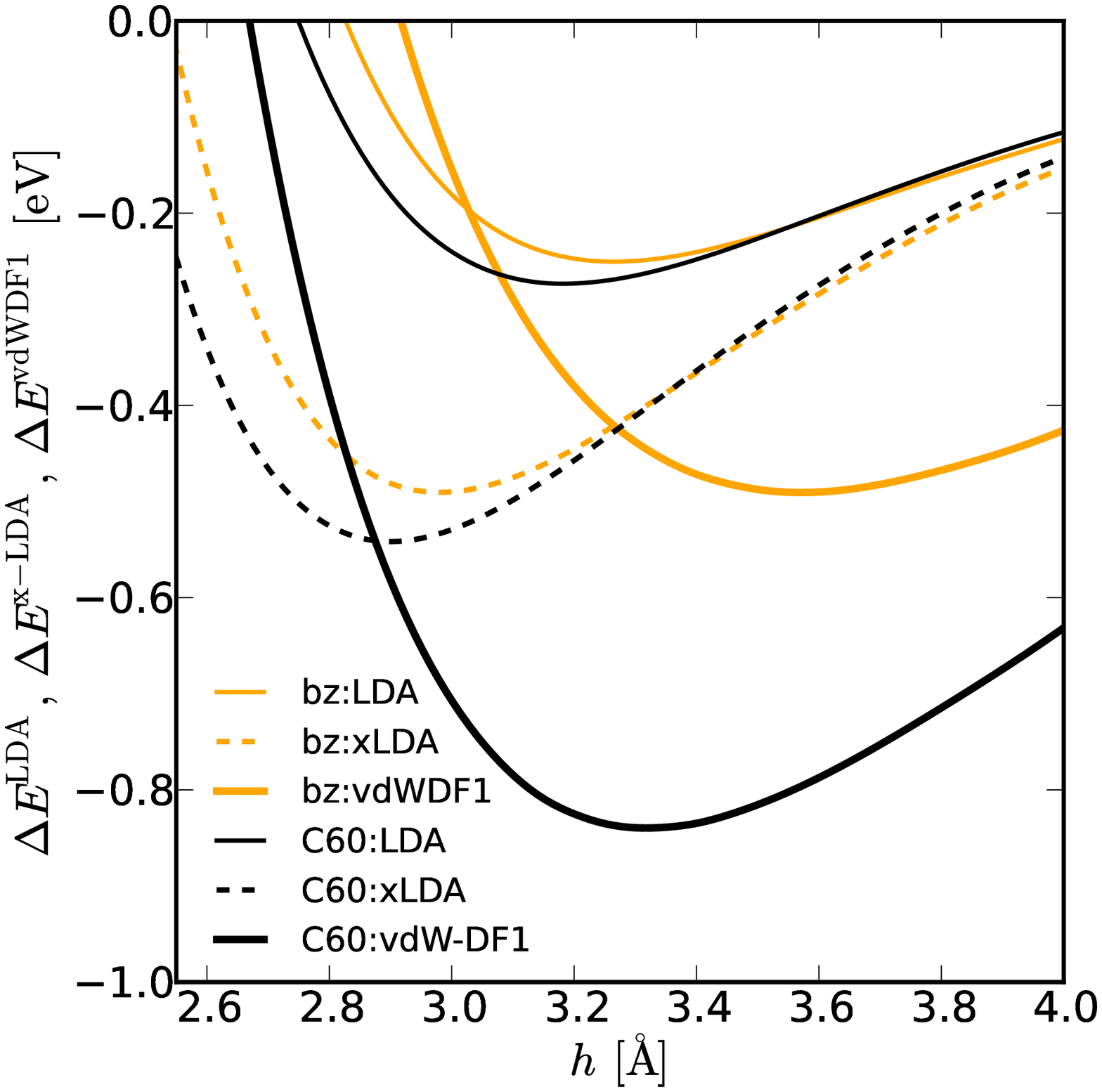}  
  \caption{Potential energy curves for benzene (bz) and C60 on graphene for LDA, a modified LDA account, and vdW-DF.}
  \label{fig:binding_curves}
  \includegraphics[width=8cm]{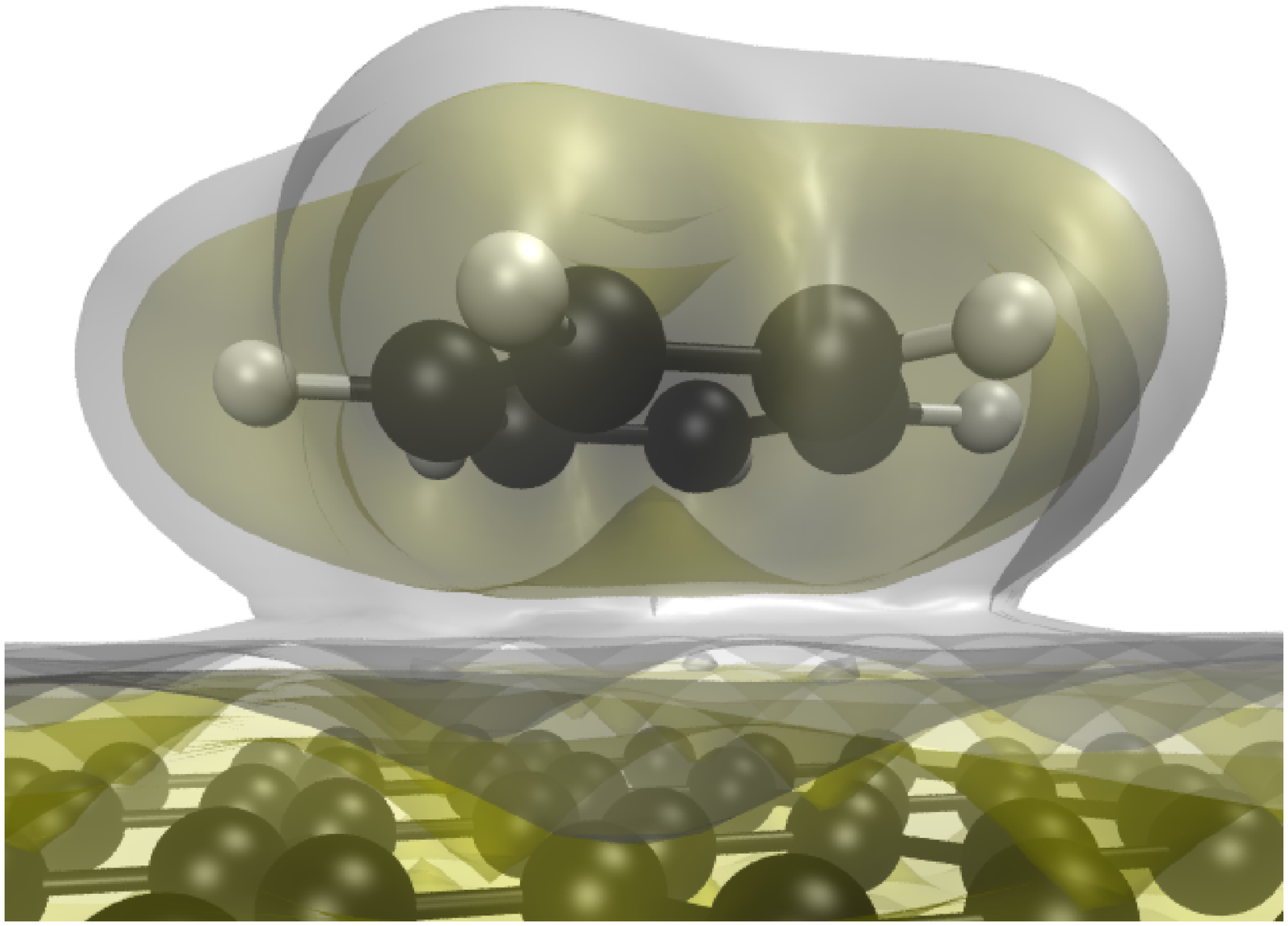}
  \caption{Density isosurfaces 
  for benzene on graphene at 3.0\,\AA. The outer isosurface is for 0.006/$a_{\rm bohr}^3$ and the inner (yellow) is for 0.027.  They encapsulate all but 1.6\% and 8\% of the electrons. 
  Figurge generated with VMD.\cite{VMD}}
  \label{fig:isosurface}
\end{figure}

The fact that LDA does not include van de Waals forces is
 well established, yet one can still find recent LDA-based studies for sparse matter. 
 Perhaps this is due to lack of awareness of DFT-D or non-local correlation functional methods. Perhaps a pragmatic point of view is taken: The underlying cause of the binding can be irrelevant for a given investigation, often the case for band structure calculations.\cite{Harris_new}

For some van der Waals bonded systems, an LDA account of  exchange-correlation may predict binding energies and separations that compare well with experimental. 
This occurs because the LDA exchange (as well as correlation) part is inadequate for sparse matter;\cite{ExcEnergy}
once replaced with a more sophisticated GGA functional,
most or all this exchange binding vanish. 
We illustrate how local approximations to exchange-correlation functional is incapable of capturing the larger binding energy of C60 compared to benzene on graphene.

Figure~\ref{fig:binding_curves} compares the potential energy curves for C60 (black full curves) and benzene (orange full curves).
The full thick curves are obtained using vdW-DF1 while the two upper thin curves are with LDA. It produces binding separations that are within the range of what the different version of vdW-DF predict. 
The binding energy is 0.25~eV, about half of the experimental. Most revealing, however, is the puny 9\% increase in LDA binding energy when comparing with C60 adsorption.

As Fig.~\ref{fig:binding_curves} confirms,
LDA does not capture the long-range van der Waals forces. Neither would of course any other local approximation to the exchange-correlation functional. 
To illustrate how a fitted local approximation might fail, we
consider the following toy exchange-correlation functional, in the spirit of $X\alpha$\cite{Slater_book},
\begin{equation} 
  E_{\rm xc}^\text{x-LDA} = (1+\alpha) E^\text{LDA}_\text{xc}\,.
  \label{eq:xLDA}
 \end{equation}
Here $\alpha$ is taken, for the sake of the argument, as a fitting parameter, which we set to $0.175$ to produce the experimental adsorption energy of benzene on graphene, with corresponding PEC given by the dashed orange curve in Fig.~\ref{fig:binding_curves}.  
The optimal separation
of about $3.0$~\AA~is likely an underestimation, but it is not that different from $3.25~$\AA, as predicted by vdW-DF1(C09). 
However, when using this fitted description for a system of different size,
the limitations, as also discussed in Ref.~\onlinecite{Rydberg:Layered}, become evident.
Applied to C60 on graphene, the same toy functional produces the black dashed curve of Fig.~\ref{fig:binding_curves}. 
The corresponding adsorption energy is merely 8\% larger than that for benzene, completely off the experimental increase of 70\%. 

Fig.~\ref{fig:isosurface} shows density isosurfaces for a benzene-to-graphene separation of 3.0\,\AA. 
The well separated inner isosurfaces encapsulate 92~\% of the total density density, while the outer, which is connected in some regions at the interface, 
encapsulate all but 1.6\% of the total density.
This small density overlap, even at such a short separation, gives a clear indication why it is futile to capture long van der Waals interactions with local density functionals, in which the binding must be described alone by the local density magnitude. 

The presented LDA analysis
also indicates the problem of
combining accounts of van der Waals forces with GGA exchange functionals that induce some exchange binding, 
in particular if fitting to unrepresentative training sets are used.
For bigger systems, in the interface-normal direction, the balance between repulsive and attractive terms can be shifted because a GGA exchange attraction only scales roughly with the size of the interface and it cannot keep up with the increased non-local attraction. 

We note that the popular PBE functional\cite{Wu_Scoles:Towards_Extending,Kannemann_Becke,ExcEnergy}  has a certain unphysical exchange attraction and that it is often used together with pair-potential accounts of van der Waals forces. 
The optPBE functional may also induce a small exchange attraction at moderate separations, and is in this sense not optimal. 
Formally, so does revPBE, but this exchange functional is so repulsive that this is not an issue in practice. 

\subsection{Role of exchange and correlation version}

\begin{figure}
  \includegraphics[width=8cm]{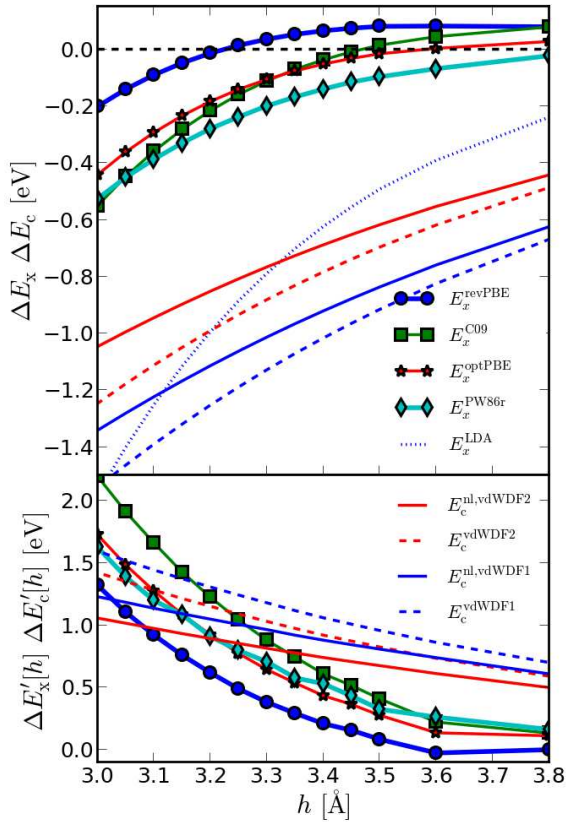}
  \caption{Upper panel: Exchange and correlation energy components to the potential energy curve~(PEC). 
  Lower panel: Derivatives PEC components. }
\label{fig:xc_bz_graphene}
\end{figure}
Figure~\ref{fig:xc_bz_graphene} presents 
the exchange and correlation components of the PEC (upper panel) and their derivates (lower panel) for benzene on graphene as obtained with different versions of vdW-DF. 
The values presented earlier in table~\ref{tab:energies} are discussed in terms of these curves. 
The upper blue curve in the upper panel (circular dots) shows the revPBE exchange part of the vdW-DF1 PEC, hereafter referred to as the revPBE curve; 
and in the same manner for the other exchange curves.  
The revPBE curve is above the other exchange curves.
Hence, in agreement with table~\ref{tab:energies}, canonical vdW-DF1 (with revPBE) gives the smallest binding energy.
The revPBE curve is also the least steep among the exchange curves, as shown by its derivative in the lower panel. This result is related to revPBE producing the largest optimal separations.

The C09 curve (green with square markers) is the steepest explaining why vdW-DF1(C09) and vdW-DF2(C09) produce the shortest separations.
The C09 curve becomes similar to the revPBE curve
at 3.8~\AA. This similarity reflects the similar enhancement factors $F_x$ of revPBE and C09 in the large $s$-limit, as mentioned in section~\ref{sec:func_version}.

The optPBE curve has about the same depth as the C09 (upper panel), but is less steep (lower panel).
Hence the adsorption energy of vdW-DF1(optPBE) is similar to that of vdW-DF1(C09), while the optimal separation is larger.

The revPBE, optPBE, and C09 curves have a positive value at a separation of 3.8~\AA.
Since these curves inevitably will decline towards zero as
the overlap between the fragments vanish, 
these exchange variants have the potential to induce spurious binding at large separations; a result related to the saturation of their enhancement factor $F_{\rm x}(s)$ at large separations.\cite{ExcEnergy}
Whether spurious binding indeed arise for a given binding curve depend on the balance between the different terms contributing to a given PEC. 

The PW86r curve stays below $0$ at all separations, thus this exchange choice does not induce binding for this system. 
For small molecular dimers, this non-binding property has been shown to be a consequence of the fact that the PW86r enhancement factor $F_{\rm x}(s)$ always increases with $s$.\cite{ExcEnergy}
Our result indicates that this non-binding property carries over to to extended systems, which further motivates choosing this exchange partner.

The full curves in the bottom part of the upper panel of Fig.~\ref{fig:xc_bz_graphene} show the non-local correlation contributions to the PEC.
The dashed include contributions from LDA correlation. 

The vdW-DF1 correlation curve is significantly deeper than the one for vdW-DF2. This result relates to vdW-DF1's larger adsorption energies.
The steepness of these two curves, shown in the lower panel, however differ much less from each other than that the steepness of the exchange curves. 
Thus, within the selected range of functional variants considered here, the binding separation is primary set by which exchange variant is chosen. 

Table~\ref{tab:energies} shows that for a given exchange variant, 
vdW-DF1 predicts merely about 0.05~\AA\,shorter binding separations than vdW-DF2  despite the much larger binding energies.
On the one hand, vdW-DF1 and vdW-DF2 correlation greatly affects the binding energy, while on the other hand, the correlation version matters less for the binding separation.
This observation makes 
the binding separation a better parameter to assess the quality of an exchange partner to vdW-DF than the energy. 
In the same vein, if a strategy of fitting exchange to high-accuracy data sets is chosen, we expect the binding separation to be the better target.

The extremely steep LDA exchange curve (dotted curve in upper pane) 
seems to explain why LDA for some systems produce binding energies comparable to a van der Waals bond but for others severely underestimate it. 
At larger separations, the curve goes rapidly to zero as the overlap between the fragments vanish. 
At short, the  curve is even deeper than the vdW-DF1 correlation curve. 
A slight shift in the kinetic-energy repulsion or electrostatic interaction could move the molecule slightly inwards or outwards, adjusting the adsorption energy greatly.

\subsection{Benzene versus C60 adsorption}

Figure~\ref{fig:bz_vs_C60} confirms 
our assumption that the reason why C60 binds more and at shorter separations than benzene arise from the increased van der Waals interaction. 
The upper curves gives $\Delta E_0^{\rm C60} - \Delta E_0^{\rm bz}$ on graphene. 
$\Delta E_0$ has a mostly-repulsive contribution the PEC of C60 and benzene.  
The lower gives the same for the non-local correlation part $\Delta E^{\rm nl,C60}_{\rm c} - \Delta E^{\rm nl,bz}_{\rm c}$. 
Since the two lower curves are far deeper than the two upper, the increased binding of C60 over benzene on graphene  arise primarily from non-local correlation. 
Similar results are found for benzene and C60 on BN. 

The figure further shows that the non-local correlation of vdW-DF1 deepens the PEC more than that of vdW-DF2 does, while the small shift from $E_0$ is similar for the two functionals. 
This result is related to the bigger role of non-local correlations in vdW-DF1. That the binding energies of the two functionals do not differ more is related to that revPBE exchange functional is more repulsive than PW86r. 
As we go from benzene to C60, the repulsive component do not change much. Therefore, the larger van der Waals component of vdW-DF1 compared to vdW-DF2, reduces the binding separation more, and increases the binding energy more.

The result that van der Waals forces is the primary cause of the increased binding of C60 compared to benzene on graphene (and BN) shows
that these systems serve as a good starting point to explore properties these forces. Results for these systems can highlight the role played by different length scales or density separations in the non-local correlation. 

\subsection{Role of different density separations}

To elucidate the role size and geometry have on the van der Waals attraction,
 we also study how much various density separations contribute to the vdW-DF interaction energy.
 We perform this analysis for benzene and C60 on graphene  and focus on the 
 non-local correlation. 
For this purpose, we introduce a lower and upper separation cutoff in the vdW-DF kernel:
\begin{equation}
  \phi_{R,\Delta }(\br_1,\br_2) =   \phi(\br_1,\br_2)  \theta[  (R+\Delta) -r_{12} ] \theta[ r_{12} - (R-\Delta )]\,,
  \label{eq:Enl_rad_theta}
\end{equation}
where $\theta$ is the Heaviside function. 
Just like for a full calculation, we subtract off, for a given kernel, the non-local correlation energy of the separate fragments (A,B),
\begin{equation}
  \Delta  E^{\rm nl}_{\rm c,R,\Delta}  =  E^{\rm nl}_{\rm c,R,\Delta}[{\rm Full}] - E^{\rm nl}_{\rm c,R,\Delta}[{\rm A}] - E^{\rm nl}_{\rm c,R,\Delta}[{\rm B}]\,.
  \label{eq:Enl_rad}
\end{equation}
This subtraction allows us to obtain a {\bf S}eparation-decomposed measure of the {\bf N}on-{\bf L}ocal correlation contribution to the {\bf I}nteraction energy (SNLI). This approach is similar in nature to the non-local correlation energy density analysis used by Lazi\'{c} and coworkers.\cite{CO:transitionMetals,graphene-boron,graphene:Ir} However, while their analysis has emphasis on which spatial regions dominate the response, ours focus on the role of density separations . 
 
\begin{figure}
  \includegraphics[width=0.5\textwidth]{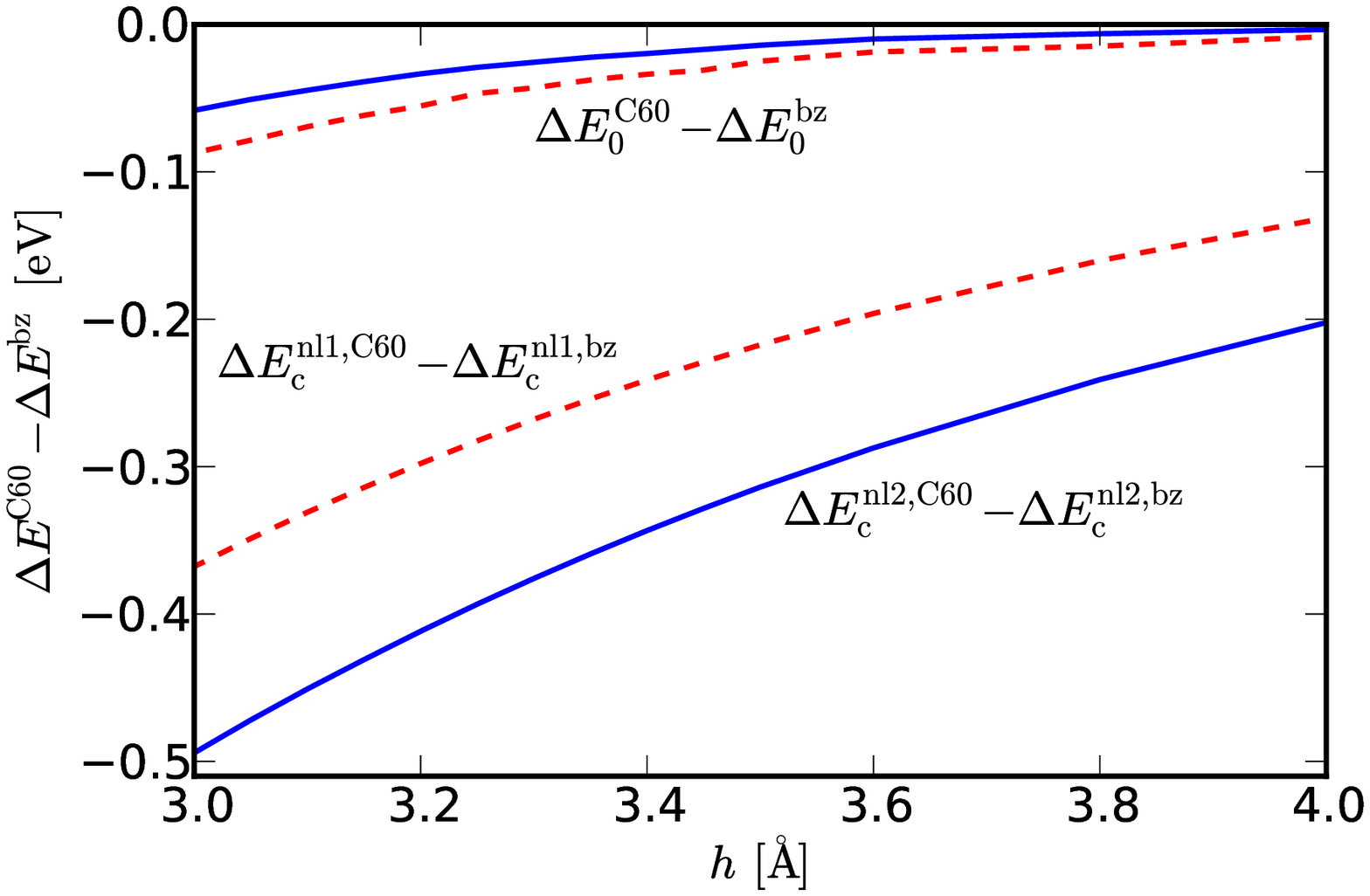}
    \caption{The difference between the components of the PEC for benzene and C60 on graphene, shown for vdW-DF1 (blue, full)) and vdW-DF2 (red,dashed).}
  \label{fig:bz_vs_C60}
  \includegraphics[width=0.5\textwidth]{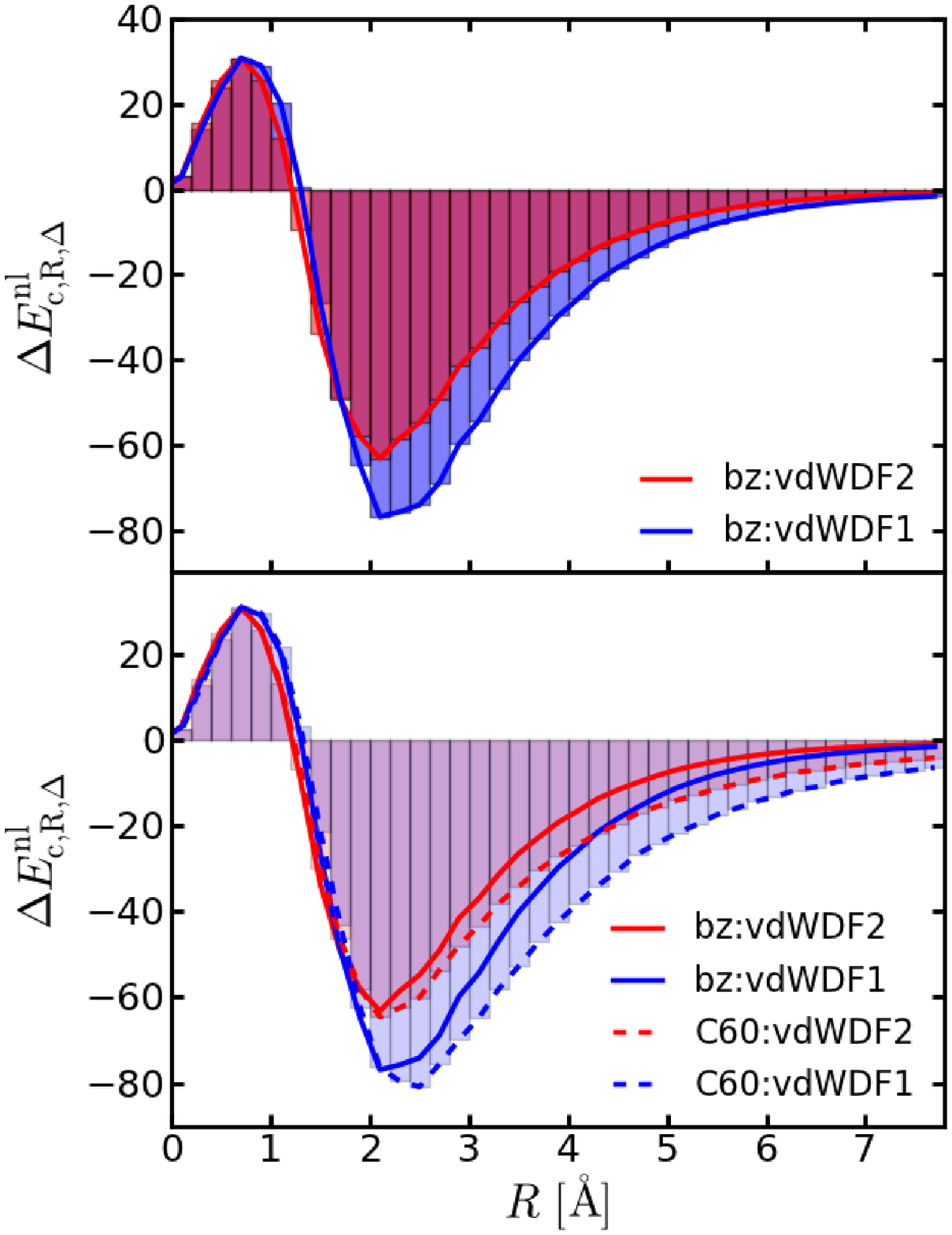}
    \caption{vdW-DF non-local correlation contributions to interaction energy
      as a function of separation between densities (SNLI) as specified by Eq.~(\ref{eq:Enl_rad}). 
        The transparent histograms show the result for benzene (bz) on graphene in the upper panel for respectively vdW-DF1 (blue) and vdW-DF2 and C60 on graphene in the lower. In both curves the full curves indicate the result for bz on graphene, while in the lower the dashed curves indicate the result for C60 on graphene.  }
  \label{fig:rad_cut1}
\end{figure}

The columns of Fig,~\ref{fig:rad_cut1} show
the SNLI for separation between $R-\Delta$ and  $R+\Delta$, with $\Delta = 0.1~{\rm \AA}$.
The upper panel is for benzene on graphene and the lower is for  C60 (and benzene once more for comparison).
The curves are guides to the eye.
In all calculations, the adsorbant has an atomic separation of $h=3.6\,\A$ from the graphene sheet, the optimal one for benzene with vdW-DF1. 
At separations smaller than about $1~\A$, the SNLI is positive, while it is negative beyond. 
Its oscillatory shape reflects the shape of vdW-DF kernel, displayed in Fig.~\ref{fig:kernel}. 
The SNLI curves in Fig.~\ref{fig:rad_cut1} are dominated by negative values because we have subtracted off the interaction within isolated fragments; 
short separations only contribute at  the interface between the two fragments. 

The SNLI of vdW-DF1 and vdW-DF2 are very similar for small density separations $R$ in these systems. 
Beyond the repulsive peak at 0.7~\A, they start to deviate, but first at about 1.7~\A\,do this deviation become substantial. 
This result is linked to the stronger asymptote of vdW-DF1 compared to vdW-DF2.\cite{H2onCu111,Vydroc:LocPol,vdW:ready} The similarity for small $R$ reflects the fact that their description of the plasmon dispersion $\omega_{\bq}$ [underlying the non-local correlation of expression~(\ref{eq:Enl})] becomes similar for large $q$.\cite{Dion:vdW,vdWDF2} 

Comparing the benzene (full) and C60 (dashed) curves in the lower panel, we see that they are almost identical for short separations: The benzene-graphene and C60-graphene interface is essentially the same as far as the non-local correlation is concerned. 
Beyond about 2~\A, the benzene and C60 curves start to differ noticeably, and at 8~\A, the contribution to the non-local correlation interaction is more than four times larger for C60 than for benzene, both in the case of vdW-DF1 and vdW-DF2. 
Thus, if we gradually consider larger systems (compared to the size of interface), the difference between the non-local correlation of vdW-DF1 and vdW-DF2 increases.
This observation is clearly also related to the stronger asymptote of vdW-DF1.

The separation analysis of Fig.~\ref{fig:rad_cut1} also highlights the strikingly different nature of a vdW-DF and a DFT-D account of van der Waals forces. 
A similar histogram for DFT-D, or strictly pair-potential account for that matter, would begin at the atomic separation, which in this case is 3.6~\AA. 
In contrast, and even despite the extended nature of the systems considered, most of the non-local correlation interaction energy in vdW-DF is already accounted for at separation shorter than 3.6~\AA.
It is not surprising that vdW-DF is enhanced at moderate separations compared to a corresponding asymptotic pair potential account as discussed in Refs.~\onlinecite{molcrys1,molcrys2}.
This enhancement is a reflection of the plasmon-nature of the vdW-DF design. 

\subsection{Analysis of density regimes}

Our analysis shows that most of the van der Waals interaction in vdW-DF arise from separations shorter than the ionic separation between the molecules. 
Because low-density regions are closer to each other, this result suggests that the low-density regions contribute the most to the interaction energy arising from the non-local correlation energy of vdW-DF. 
To quantify the role of low-density regions, we 
introduce a low-density cutoff in evaluating the non-local correlation of Eq.~(\ref{eq:Enl}) (rather than separation cutoffs). 
By gradually increasing this cutoff, we systematically  gauge the role of various-density regions. 

\begin{figure}[t]
  \centering
  \includegraphics[width=0.5\textwidth]{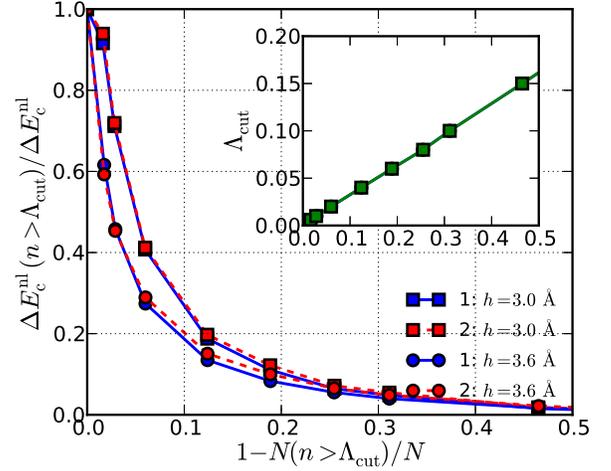}
  \caption{Fraction of non-local correlation interaction that remains as a low-density cutoff is introduced, expressed as a function of the fraction electrons removed. 
  The blue (red) square dots is for benzene at a molecule-to-sheet separation of $h = 3.0~$\AA~using vdW-DF1 (vdW-DF2), while the circular dots indicates the same for a separation of $h = 3.6$~\AA. 
  The insert relates the low-density cutoff $\Lambda_{\rm cut}$ (in terms of atomic units) to the fraction of density removed. }
  \label{fig:dens_analysis}
\end{figure} 
Figure~\ref{fig:dens_analysis} shows the result of this density analysis for benzene on graphene at two representative separation of $h = 3.0~$\AA\, and $h = 3.6$~\AA.
The vertical axis indicates the fraction of 
the non-local correlation interaction energy $\Delta E^{\rm nl}_{\rm c}$ that remains after a given low-density cutoff has been introduced. 
The same density cutoff is used in the full calculation as in the reference calculations. 
The horizontal axis indicates the fraction of the total number of electrons that has been removed in the full system for a given density cutoff.
The insert relates the density cutoff [in atomic units] (vertical axis) to the fraction of electrons removed (horizontal axis).

The figure reveals that the vdW-DF interaction energy is very sensitive to low-density regions. 
For the case of $h=$3.0~\AA, removing 2\% of the density (as indicated in the outer contour in Fig.~\ref{fig:isosurface}) removes 15\% of the interaction energy; 
removing 8~\% of the density (indicated in the inner counter of Fig.~\ref{fig:isosurface}) removes 65\% of the interaction energy. If we remove 50~\%~of the density, starting from the smallest density, 97\% of the interaction energy is removed. For a vdW-DF like functional that weighted density equally, the non-local correlation would be quadratic in the density and  the corresponding numbers would be 4\%, 15\%, and 75\%. These widely different trends reflects how the kernel $\phi[n]$ itself is a functional of the density $n$.

Low density regions are even more important at the larger separation $h=3.6~$\AA.
To crudely explain this effect, we can consider that when the molecules move further apart, the density between the molecules is lowered, because the density overlap of the molecules is reduced.
 At the larger separations, we also find that vdW-DF1 is somewhat more sensitive to low density regions than vdW-DF2, while at 3.0~\AA~the curves are very similar.
This trend agrees well with the separation analysis shown Fig.~\ref{fig:rad_cut1}, since vdW-DF1 and vdW-DF2 differ more at larger separations.

Our analysis show that interactions between different low density regions (as connected by the two-point kernel $\phi(\br,\br')$)  dominate the contributions to the interaction energy. 
Clearly the response of low-density regions is much stronger that that of high density regions. 
Since part of the low-density regions are closer to each other (the region close to the interface) this result can be linked to the importance of shorter density separations (than atomic separations between fragments) established in the previous subsection. 

Finally, we note that the importance of low-density regions has an important bearing on how we build our intuition of how strong a van der Waals interaction will be.
According to the presented analysis, it is not the number of atoms that roughly scale with strength of van der Waals interactions but more the amount of low-density regions in a fragment. This observation can be for instance linked to the offset in the binding trends of PAHs\cite{svetla:PAH,Bjork_aromatic} and alkanes\cite{Londero:alkanes} on graphene and the big response of hydrogen atoms.\cite{molcrys1,Bjork_aromatic}

\begin{figure*}[t!]
  \centering
  \includegraphics[width=1.0\textwidth]{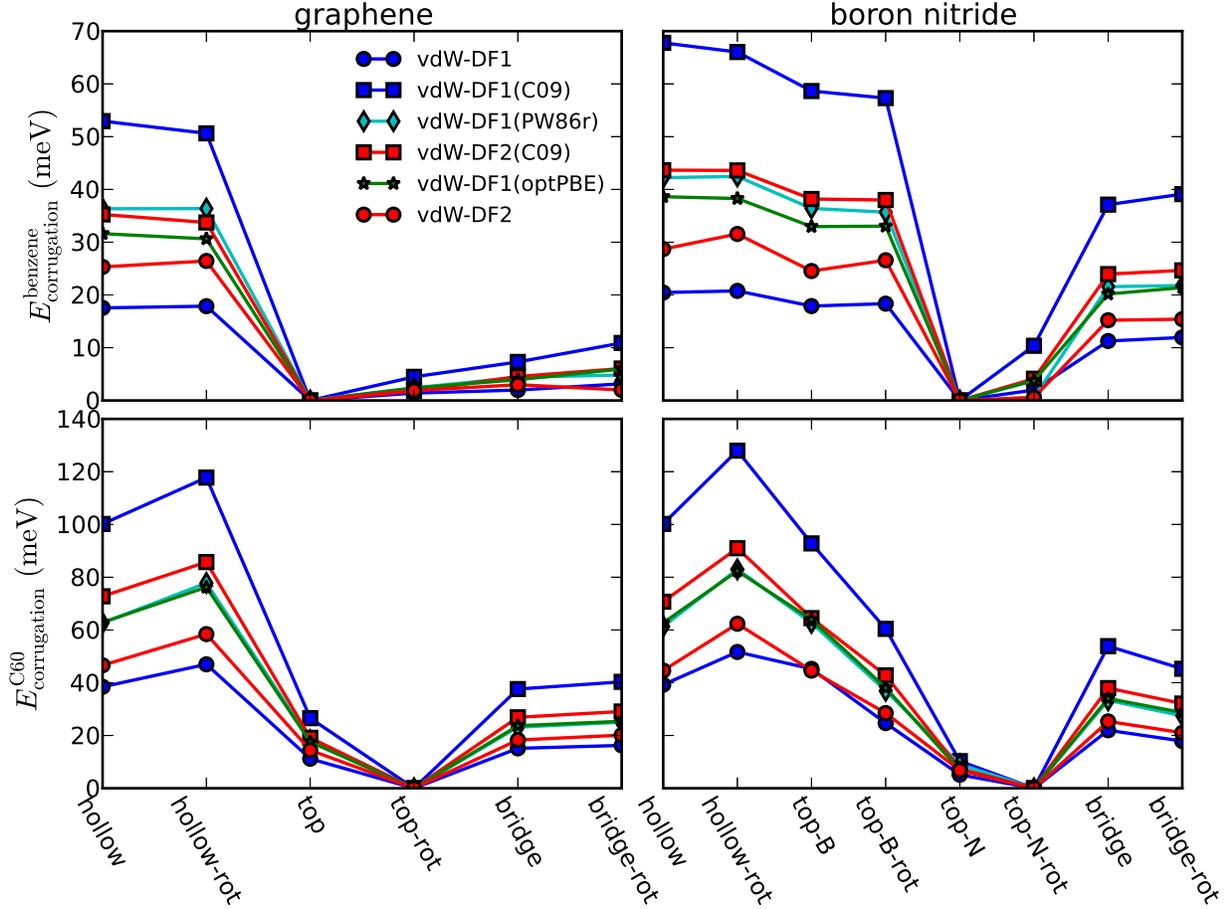}  
  \caption{The variation in adsorption energy as a function of the site, 
    for benzene (upper row) and C60 (lower row) adsorbants  on graphene (left column) and on boron nitride (right column). The data points (connected by lines) are obtained with different variants of vdW-DF. }
  \label{fig:corr2}
\end{figure*}

\section{Corrugation results}
\label{sec:corrugation}

The predicted corrugation is very sensitive to which version of vdW-DF and exchange variant is used. 
Corrugation is here understood as the potential energy variation of a molecule on different sites on the surface. We have recently shown\cite{H2onCu111_B} for H$_2$ on Cu(111) that the corrugation is even more sensitive to switching between a vdW-DF to a DFT-D account.
The adsorption energies and optimal separations, presented in section~\ref{sec:Adsorption}, 
is also significantly affected by the vdW-DF variant used; for benzene on graphene, the optimal separations differ by $8\%$ and the energy by $70\%$. 
Nevertheless, the vdW-DF sensitivity to corrugation stands out: for this system, the diffusion barriers can change by up to a factor of three as we switch the exchange variant and correlation version. 

This huge sensitivity arise because the corrugation depends exponentially on the binding separation; slight changes in separation greatly affect the corrugation.

Figure~\ref{fig:corr2} shows the how the optimal binding energies varies for the different high-symmetry sites on graphene and BN for benzene and C60 adsorbants, as calculated with the considerent variants of vdW-DF. 
It shows that the corrugation depends strongly on the correlation version used and in particular on the exchange choice. 
The upper  panels show the results for benzene on graphene and on BN. The lower, show those for C60 with the carbon hexagon facing graphene and BN, which is preferred over the pentagon face.
Graphene sites are identified in Fig.~\ref{fig:site_bz_g};
the BN sites are similar, but with twice as many top sites: the  hexagon can be centered on N or on B (top-N and top-B). 
The vertical scale in the lower panels are twice those of the upper; the corrugation of C60 adsorbants is roughly twice that of benzene. 

Benzene can easily diffuse on graphene since the molecule can pass between top and bridge sites which have very similar energies. 
This open-path dynamics is similar to that of benzene on Cu(111).\cite{benzCu,H2onCu111_B}
On BN this path is closed, since from a top-N site, the molecule must either move to a top-B site or a hollow (or between) to get to the next top-N site.  
The corrugation energies are also generally somewhat larger. 
Thus our calculations indicate that benzene will freeze at much higher temperatures on BN than on graphene. 
C60 has a similar, though less pronounced, low-energy path on 
graphene. 
Rotational barriers are in general small both for benzene and C60. 
We note that since C60 adsorbants induce a dipole in particular on graphene, in-plane dynamics must be accompanied by continuous charge redistribution. Since electron-hole excitations express such charge-relocations, this will open a damping mechanism causing friction\cite{BPersson:Damping,PerssonHelssing:Damping,PerssonPersson} for the dynamics for C60 on graphene. 

There is a question of whether C60 might diffuse like a soccer ball on graphene. According to classical molecular dynamics calculations, this is not the case for thermally induced motion.\cite{Diffusive_C60} However, such a classical analysis fail to account for dynamics of the friction that we argue will be present. 

Comparing the functionals, we find that 
vdW-DF1 predicts the smallest corrugations followed by vdW-DF2, while vdW-DF1(C09) clearly predicts the largest. In between are vdW-DF2(C09), vdW-DF1(PW86r), and vdW-DF1(optPBE) which are more or less tied. 
The succession follows closely that of the binding separations from smallest to largest as listed in table~\ref{tab:energies}.
For benzene, vdW-DF1 clearly predicts a larger corrugation than vdW-DF2, but for C60 the difference is smaller. This trend is reflected in the more similar separations predicted for C60 adsorption.  

\begin{figure}[t]
  \centering
  \includegraphics[width=0.48\textwidth]{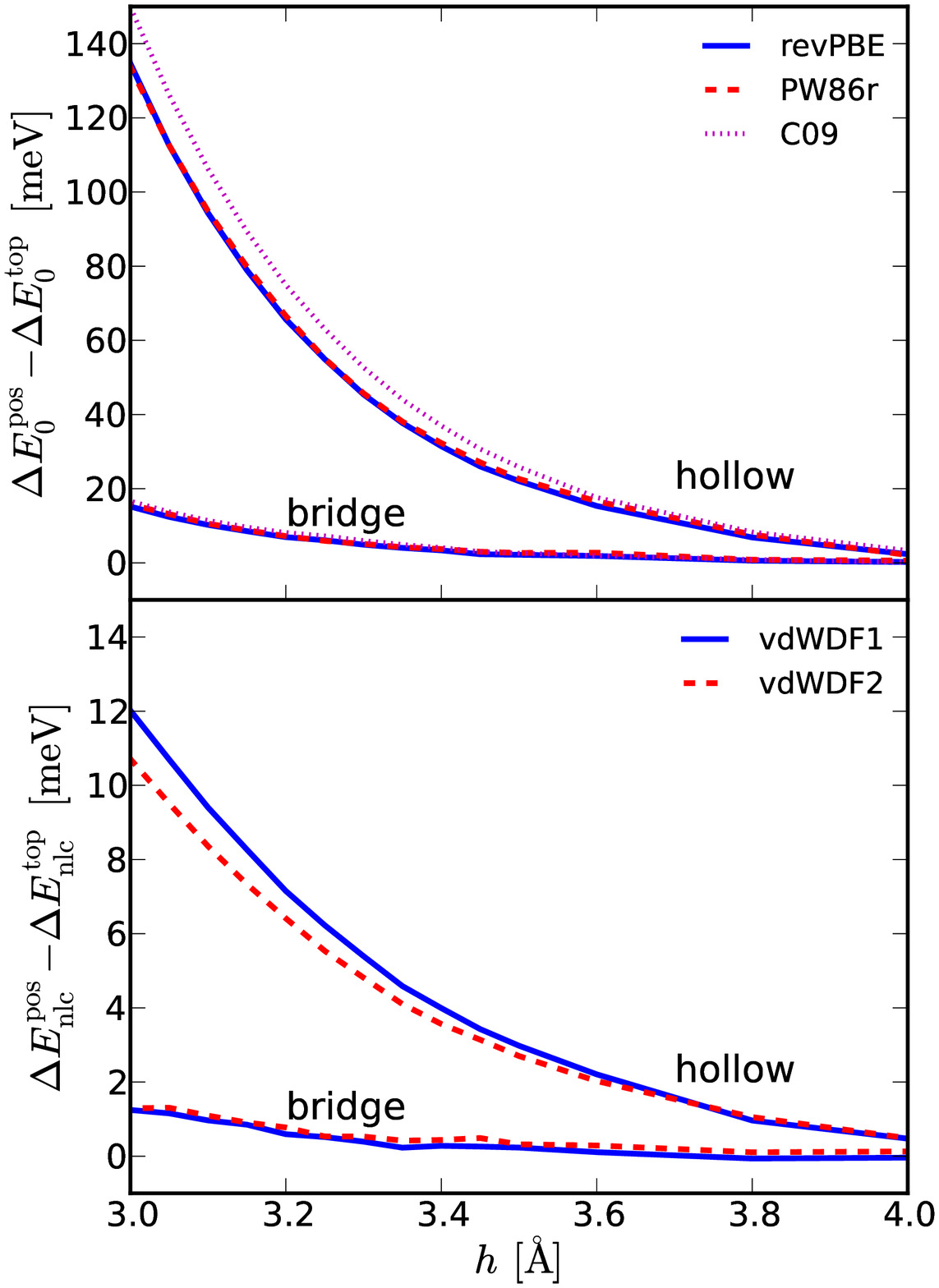}  
  \caption{Corrugation sensitivity to separation $h$ between benzene adsorbant and graphene sheet.   Upper panel: Difference between $\Delta E_0(h)$ for a given site and the top site for three different exchange functionals. 
  Lower panel: In the same manner for the non-local correlation part $\Delta E^{\rm nl}_{\rm c}$ for vdW-DF1 and vdW-DF2.   }
  \label{fig:corr_analysis}
\end{figure}

The observation of a strong link between optimal separation and corrugation is 
 somewhat clouded by fact that the separation is optimized separately for each site.
To make a clearer analysis, we consider how the difference between the PECs of two specific sites ($\Delta$PEC) changes if we change the exchange variant or correlation version. 

Fig.~\ref{fig:corr_analysis} shows, for benzene on graphene, that $\Delta$PEC barely changes
with the exchange variant and correlation version used.
In contrast, the full PEC might change much.
Thus, the corrugation of graphene and BN is predominantly set by the binding separation (given by the minimum of the full PEC), making the huge sensitivity to functional version arise indirectly as consequence of the exponential sensitivity to binding separation. 

In Fig.~\ref{fig:corr_analysis}, the indirect sensitivity to corrugation is documented separately for the mostly-repulsive $\Delta E_0$ part of the PEC (upper panel) and for the non-local correlation part $\Delta E^{\rm nl}_{\rm c}$ (lower panel). 
In the upper panel, the difference between $\Delta E_0$ of the top site and that of the bridge and hollow, for the exchange choice of revPBE, PW86r, and C09. And in the same manner in the lower panel for $\Delta E^{\rm nl}_{\rm c}$ for vdW-DF1 and vdW-DF2 correlation.
The curves conform closely to each other (the same goes for optPBE exchange which is not shown).
The  small difference in the $\Delta E_0$ curve is much bigger that the difference between the non-local correlation contribution curves $\Delta E^{\rm nl}_{\rm c}$.  
The vertical axis of the lower panel spans one-tenth of that of the upper panel.

The similarity of the tiny contributions to corrugation arising directly from vdW-DF1 and vdW-DF2 correlation can be explained by the similarity of their SNLI curves shown in Fig.~\ref{fig:rad_cut1} for short separations.
The curves in the upper panel depend essentially exponentially
on separation --- for the largest separations noise contributes to the tiny difference between the top and hollow PEC. 
The curves document how corrugation is set by the optimal separation. It is the effect the exchange-correlation variant has on the binding separation that causes the corrugation to be strongly affected by the variant.

\section{Towards a surface-based experimental reference database}
\label{sec:discussion}

Adsorption on clean surfaces provide outstanding opportunities to compare theory and experiments. 
Specific molecules can be carefully deposited on surfaces\cite{Bartels:Tailoring} and imaged with scanning-tunneling (STM)  or atomic-force microscopy (AFM). 
Several experimental probes can be used to gain detailed quantitative insight that can be compared with calculated numbers.
One example is the comparison of theoretical and experimental vacuum-level shifts that can be used to deduce binding separations and charge transfer of weakly chemisorbed molecules.\cite{others:benzCu,Toyoda:pentacene,Toyoda:Perfluoropentacene}
Another is the measurements of resonance levels arising in back-scattering of light molecules, which provide a powerful 
connection between experiments and theory.\cite{SAndersson:H2Cu_PRL,SAndersoon:Cu100,SAndersson:H2Cu,HeavyAtoms,H2onCu111,H2onCu111_B}
These experiments can be used to deduce the shape of the PECs as well as corrugation and could give detailed information on interactions at many different length scales. 

\subsection{Proposed concerted theory-experiment approach}

Based on our results and analysis, we propose a concerted
theory-experiment approach to 
probe the magnitude of the various exchange-correlation contributions to physisorption on surfaces.
The proposed approach is complimentary to others that emphasize a back-scattering based comparison\cite{SAndersson:H2Cu_PRL,SAndersoon:Cu100,SAndersson:H2Cu,HeavyAtoms,H2onCu111,H2onCu111_B}
 (limited to light molecules) or vacuum-level shift comparison\cite{others:benzCu,Toyoda:pentacene,Toyoda:Perfluoropentacene}
 (limited to systems with a sizeable charge transfer at the interface). 

The suggested approach takes three steps:
First, one compares experimental and calculated corrugation and thus determine the binding separation. 
Second, based on the separation, one estimates how much of the adsorption energy arise directly from the exchange-correlation energy.
Third, comparing the adsorption of  molecules with the same interface geometry but different shapes and volume, like benzene and C60, one probes how much of the adsorption energy arise from the van der Waals interaction.

{\it The first step} is motivated by the fact that the predicted corrugations are strongly sensitive to the vdW-DF variant ---in particular to the exchange (Fig.~\ref{fig:corr2})---
while at the same time the difference between PECs of two sites is very insensitive to the variant used~(Fig.~\ref{fig:corr_analysis}).
This result indicates that the strong corrugation sensitivity, at least for benzene and C60 adsorption on BN and graphene, arise almost exclusively from the
shift in separation induced by the replacement of exchange-correlation functional.
Turning this around, we can can conclude that 
the corrugation essentially determines the separation. 
If the corrugation can be measured, DFT calculations can determine the binding separation, since the specific exchange-correlation used for this comparison is unimportant.

Many other systems might also exhibit this kind of direct exchange-correlation insensitivity.
The identification of additional systems of this kind merits further study, in particular if the corrugation has been accurately measured. 
The validity of the argument presented here is also limited by the fact that the exchange-correlation account affects the Kohn-Sham (KS) orbitals and thereby kinetic energy contributions. 
However, for purely van der Waals bonded systems, the direct part of this effect is likely small.

{\it The second step} also builds on the assumption that the Kohn-Sham orbitals are not much affected by the specific exchange-correlation functional used. This assumption is appropriate for weakly-bonded systems.\cite{Harris_new}
If so, the total exchange-correlation contribution to the adsorption energy can be assessed if both the binding separation and energy is known. 
The separation also tells us how the exchange-correlation functional affects the steepness of the PEC, since the van der Waals forces must counter the kinetic-energy repulsion to make the molecule physisorb at a given separation. 

{\it The third step} builds on the result that for molecules with
similar interfaces to the surface (or sheet) like benzene and C60, but with different geometries and size, we can single out the effect of van der Waals forces (Fig.~\ref{fig:bz_vs_C60}). 
Such results give us insight on the delicate balance between the exchange and correlation contributions to the binding.
These results can be combined with 
results for the asymptotic interactions, since advanced methods can compute these to high accuracy.\cite{Kumar:C6_benzene,chu:C6,Unified,spackman:C6,Ruzsinszky:C60,Perdew:C60,Dobson:RPA_asymptotic}
Such analysis tells us how the van der Waals forces are modulated at shorter separations compared to the asymptotic form.\cite{molcrys1,molcrys2,Perdew:Asymptotic,Dobson:DispersionEnergies,Dobson:_asymptotic}

How to accurately determine the corrugation experimentally is a question beyond the scope of this paper, but we note that the advancements in STM and AFM is astonishing. 
One might drag molecules across the surface and measure the potential energy landscape,\cite{Force:MoveAtom,Force:MoveMolecule} or observe the diffusion of molecules on the surface.\cite{wong:Dynamics,OrganicDiffusion:Pt111}
 Based on diffusion barriers, corrugation can be estimated.

\subsection{Surface-based database}

Many adsorption systems may have the structural relation as benzene and C60 on graphene and BN and that could make them good discriminators of exchange-correlation accounts. 
Other experimental-systems, like scattering of light molecules on noble metal surfaces or systems with vacuum level shifts also provide excellent links between theory and experiment. 
As experimental methods continue to develop and new experimental data becomes available, these could be linked to DFT calculations or parameterized in such a manner that it becomes straightforward to assess the performance of sparse matter methods.  
The  construction of PEC based on resonance levels of H$_2$ backscattering\cite{SAndersson:H2Cu_PRL,SAndersson:H2Cu} makes it simple to assess the theoretical methods.\cite{H2onCu111,H2onCu111_B}   The relation between corrugation and binding separation discussed in section~\ref{sec:corrugation} provides a different example. 

If we identify and organize several of these theory-experiment links for surfaces  we can build up a database that could provide invaluable for the continued advancement of sparse-matter methods.
Today datasets based on high-accuracy quantum-chemistry calculations play such a role,\cite{vdW:ready,S22,S22:5,S66} but surface electrons have a different collective nature than that of small molecular dimers.\cite{HR_thesis,Lan:vdWApp}
Surfaces are also especially suited to extract a range of experimental properties. 
The corrugation sensitivity to binding separation is central to the proposed approach. Other quantities sensitive to geometry, like the HOMO-LUMO gap of adsorbed molecules,\cite{Kubatkin,Neaton:ReNormalization}
may help uncover other links between theory and experiment useful for assessing sparse-matter methods. 

\section{Summary}
\label{sec:conclusion}

In this paper, we identify a comparison of the adsorption of benzene and C60 on inert surfaces like graphene and BN as a good arena to evaluate exchange-correlation accounts of van der Waals forces and to gain insight into the nature of physisorption.
Our study of benzene and C60 on graphene have shown that vdW-DF1 and vdW-DF2 accurately capture the increased van der Waals binding of the larger molecule.

A detailed analysis of adsorption results, exchange-correlation components to the PECs, and corrugation have been performed.
Based on this analysis, we have made a number of observations.
We have shown that C60 induces a moderate dipole on graphene, while for the other systems dipoles are tiny.
We have shown that the binding separation is more sensitive to the exchange variant than the correlation version of vdW-DF.
This result making the separation a better target for assessing or optimizing an exchange account. 
We have shown that the difference between benzene and C60 binding arise predominantly from the increased non-local correlation, displaying fundamental shortcomings of relying on unphysical exchange binding.
With our density separation decomposition of the non-local correlation interaction, we documented that vdW-DF1 and vdW-DF2 are similar for short density separations, but differ much at large.
The similar density-cutoff analysis reveals how the non-local correlation of vdW-DF is very sensitive to low-density regions.

Further, we have demonstrated that the corrugation is very sensitive to the vdW-DF variant used, but also that a particular low-energy path exists for benzene on graphene, making it easy for this molecule to diffuse. The corrugation sensitivity have been shown to arise almost exclusively as an indirect effect arising from the shift in binding separation caused by the exchange choice and correlation versions.
This result has lead us to suggest a experiment-theory approach to assess variants of, and gain insight into, the exchange-correlation functional and its account of van der Waals forces.

The novel analysis approaches presented here, like density-separation  and density-regime decomposition,
can also be applied to other material systems and functionals. Such approaches can allow us to view materials and methods in new ways helping the development of non-local correlation functionals

\acknowledgments
We thank T.~L\"ofwander and B.~I.~Lundqvist for discussions. 
The work was supported by the Swedish Research Council (VR) and the Chalmers Area of Advance.
The Swedish National Infrastructure for Computing (SNIC) at the C3SE and HPC2N is acknowledged for computer time.

\end{document}